\documentclass[review]{elsarticle}

\usepackage{lineno,hyperref}
\modulolinenumbers[5]

\journal{Journal of Solid State Electronics}

\usepackage{subcaption}
\usepackage{amsmath}
\usepackage{array}
\usepackage{tabularx,ragged2e,booktabs,caption}
\begin{document}

\begin{frontmatter}

\title{Analytical Modeling of Metal Gate Granularity based Threshold Voltage
Variability in NWFET \tnoteref{mytitlenote}}

\author{P Harsha Vardhan\fnref{myfootnote}}
\author{Sushant Mittal}
\author{Swaroop Ganguly}
\author{Udayan Ganguly\fnref{myfootnote}}
\address{Department of Electrical Engineering, Indian Institute of Technology Bombay, Powai, Mumbai-400076, India}
\address{email: phv@ee.iitb.ac.in; udayan@ee.iitb.ac.in}

\begin{abstract}
Estimation of threshold voltage $V_T$ variability for
NWFETs has been computationally expensive due to lack of
analytical models. Variability estimation of NWFET is essential to
design the next generation logic circuits. Compared to any other
process induced variabilities, Metal Gate Granularity (MGG) is of
paramount importance due to its large impact on $V_T$ variability.
Here, an analytical model is proposed to estimate $V_T$ variability
caused by MGG. We extend our earlier FinFET based MGG
model to a cylindrical NWFET by satisfying three additional
requirements. First, the gate dielectric layer is replaced by Silicon
of electro-statically equivalent thickness using long cylinder
approximation; Second, metal grains in NWFETs satisfy periodic
boundary condition in azimuthal direction; Third, electrostatics is
analytically solved in cylindrical polar coordinates with gate
boundary condition defined by MGG. We show that quantum
effects only shift the mean of the $V_T$ distribution without
significant impact on the variability estimated by our
electrostatics-based model. The $V_T$ distribution estimated by our
model matches TCAD simulations. The model quantitatively
captures grain size dependence with $\sigma(V_T)$ with excellent
accuracy ($6\%$error) compared to stochastic 3D TCAD
simulations, which is a significant improvement over the state-of-
the-art model with fails to produce even a qualitative agreement.
The proposed model is $63\times$ faster compared to commercial TCAD
simulations.
\end{abstract}

\end{frontmatter}

\linenumbers

\section{Introduction}

NWFET (Nanowire Field Effect Transistor) has emerged as the potential replacement for FinFET as an ultimate scaled CMOS device \cite{ref1}. For both FinFET and NWFET, process
induced variabilities have been very challenging from circuit performance viewpoint. Multi-gate transistors exhibit higher Metal Gate Granularity (MGG) variability compared other variability phenomena like Line Edge Roughness (LER), Random Dopant Fluctuation (RDF) etc., \cite{ref2,ref3,ref4,ref5}. The wider the $V_T$ distribution, the worse is the circuit performance \cite{ref6}. Hence, the estimation of $V_T$ distribution is a must before circuit design.
\begin{figure}[h]
\centering
\includegraphics[scale=0.2]{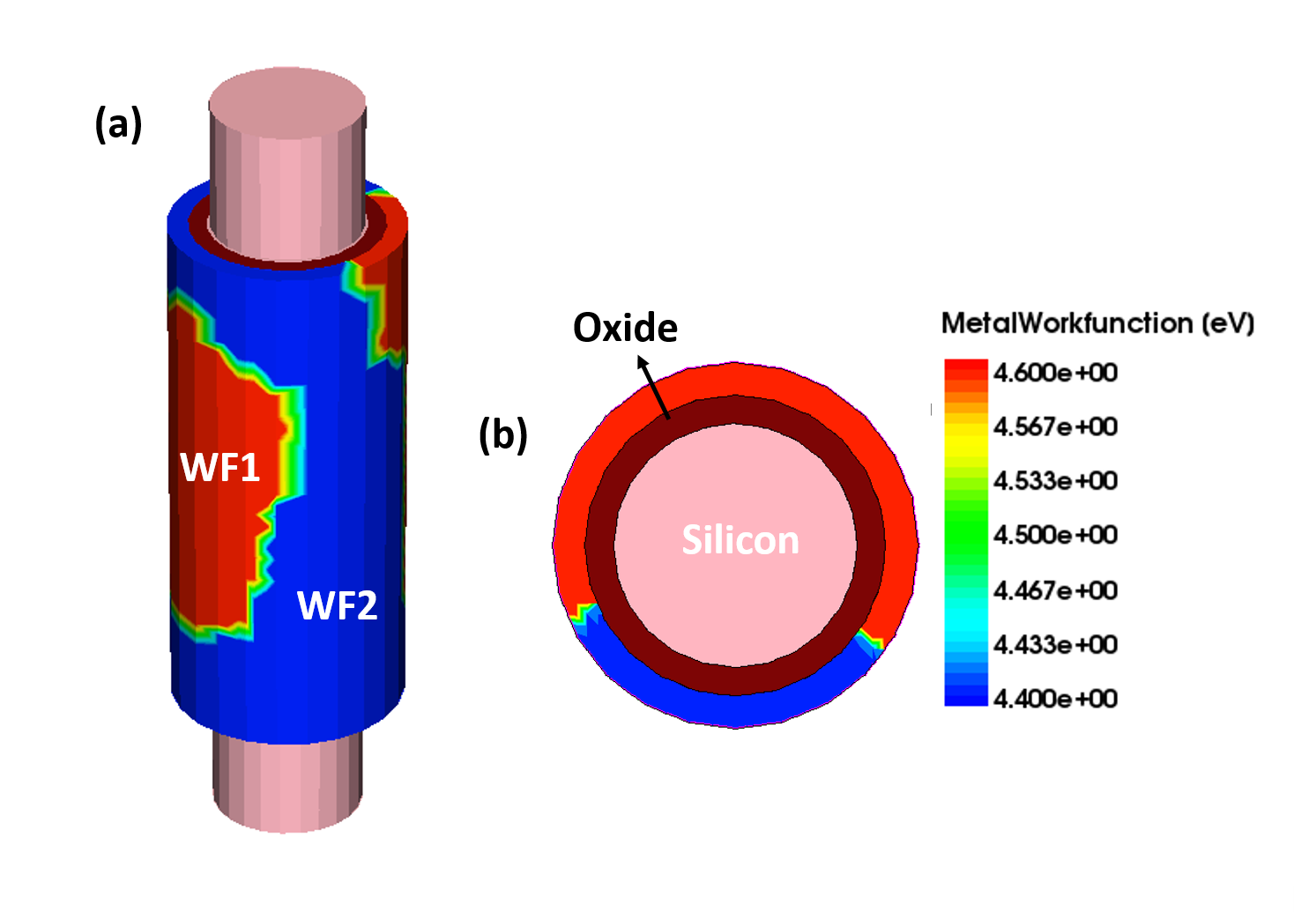}
\caption{(a) Structure of the NWFET with MGG which constitutes two
distinct work-function on the Gate. The red and blue regions show two
types of work functions (b) Cross-sectional view of the NWFET with
gate metal, Silicon (channel) and oxide.}
\label{fig:motivation}
\end{figure}


It was reported that the NWFET (gate all around nanowire MOSFET) has better immunity against MGG induced variability compared to that of FinFET by 3D TCAD simulations in \cite{ref7}. The smaller $V_T$ variability in NWFET is a strong motivation to replace FinFET with NWFET in future technology nodes. The steep computation cost of 3D stochastic TCAD simulations to extract variability seeks for more computationally efficient techniques. The $V_T$ distribution from smaller grain sizes typically result in a Gaussian like distribution while the larger grains produce a bimodal distribution (in case of two WF values). Any analytical method or model should be able to capture this size dependence efficiently. Further, analytical models are intuitive, faster than TCAD and can be a stepping-stone to a compact model. Recently, our group has developed the analytical estimation of the $V_T$ variability in FinFETs \cite{ref8}.

In this work, we extend the FinFET based MGG model to NWFET with cylindrical geometry by satisfying three
additional constraints. The results from this model agree well with TCAD. Quantum confinement (not accounted in our
model) produces a shift in the $V_T$ distribution without affecting its shape. The model is compared to state-of-the-art to show significant improvement achieved.

\section{The Analytical Model}

The body potential of NWFET is analytically modeled in \cite{ref10,ref11}. Due to very low free carrier concentration in sub-threshold regime, the electrostatics of the NWFET can be captured by the Laplace equation alone instead of Poisson equation \cite{ref10}. A 3D schematic of the NWFET with MGG is shown in Fig.1(a). A simple cylinder is assumed for the purpose of solving the Laplace equation.
\begin{figure}[h]
\centering
\includegraphics[scale=0.25]{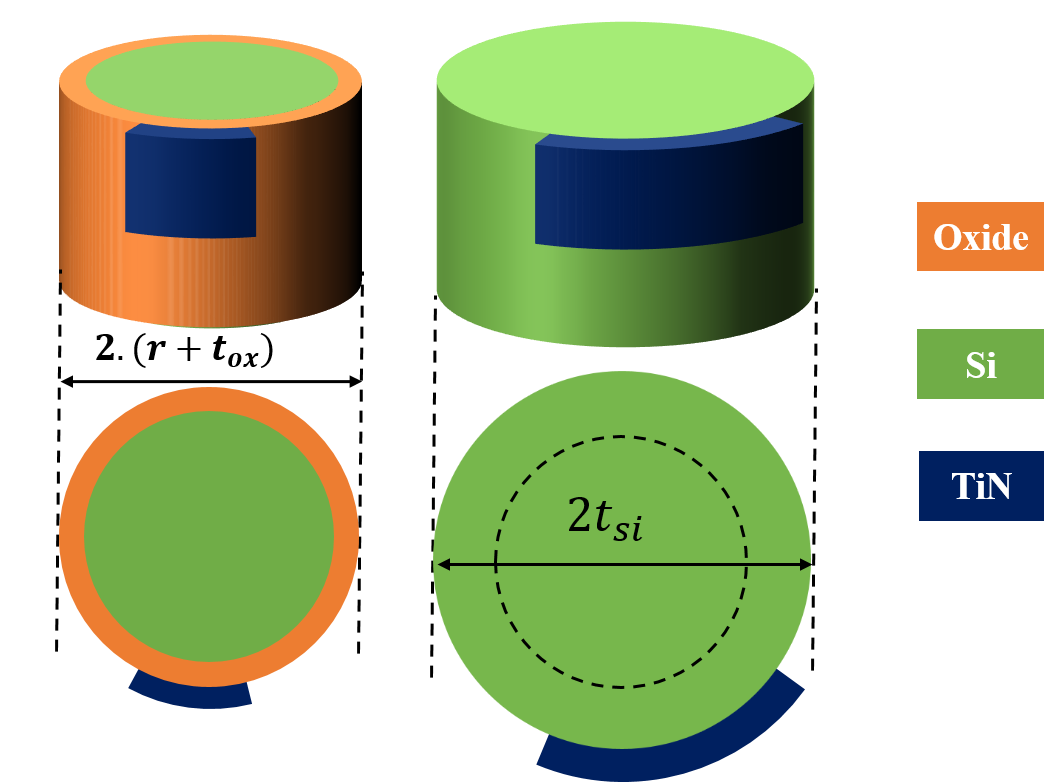}
\caption{Gate oxide on the Silicon nanowire is replaced by electrostatic
equivalent thick silicon. The size of the grain is elongated by the factor ‘$\eta$’
(only in azimuthal direction) due to this transformation.}
\label{fig:transform}
\end{figure}
Two flat faces (planes) of the cylinder correspond to two boundary conditions (Source and the Drain) and the curved surface of the cylinder corresponds to the third boundary condition i.e. the Gate. The thin gate dielectric is substituted with electro-statically equivalent, thicker Silicon to form a single homogeneous cylinder. This simplification is essential for obtaining a closed form expression for the electrostatic potential. Due to this, the radius gets modified to effective (new) radius ($t_{si}$) and is given by \eqref{eq:effectiveSit}.
\begin{equation} \label{eq:effectiveSit}
t_{si} = r \times e^{\frac{\epsilon_{si}}{\epsilon_{ox}}\ln(\frac{r+t_{ox}}{r})}
\end{equation}
where $r(=\frac{d}{2})$ is the radius of the nanowire, $d$ is the diameter, $\epsilon_{si}$ and $\epsilon_{ox}$ are the relative permittivities of the Silicon
and the Oxide. The Gate length ($L_g$) remains unchanged. Due to this transformation, the actual grain gets stretched out only in the horizontal direction (azimuthal direction) by the factor $\eta$ as given by \eqref{eq:effectiveSit}. This grain distortion is illustrated in Fig. 2. The procedure of finding variability in $V_T$ distribution is presented in three consecutive subsections below.

\subsection{Cylindrical Gate MGG Generation}
In this section, the algorithm for generating random metal gate grains is presented. This is similar to the algorithm used for FinFET in \cite{ref8} except a grain periodicity constraint in the azimuthal direction. Since the gate is wrapped around the nanowire, metal grains on the NWFET gate satisfies the periodic boundary condition in azimuthal ($\phi$) direction for the Potential ($V$). The modified algorithm is presented below. The average grain diameter ($S$) is assumed to vary from 3-25 nm \cite{ref9}. The cut-open view of the gate is shown in Fig. 3(a).
\begin{enumerate}
\item The mean number of grains ($N_G$) on the gate of a single NWFET is determined using $N_G=\frac{4 L_g \times(2\pi t_{si})}{\pi \eta S^2}$
\item The total gate area is discretized into a fine square mesh as shown in Fig. 3(a). Using ‘$N_G$’ as the mean and standard deviation of the Poisson distribution,
$N_x \times N_y$ (assuming $N_x$ and $N_y$ is the grid size in $x$ and $y$ directions) random numbers are produced.
\item These randomly generated integers are then assigned to every grid-point. The location of a non-zero integer is marked as grain center (red point) as seen in Fig.
3(a).
\begin{figure}[h]
\centering
\includegraphics[scale=0.23]{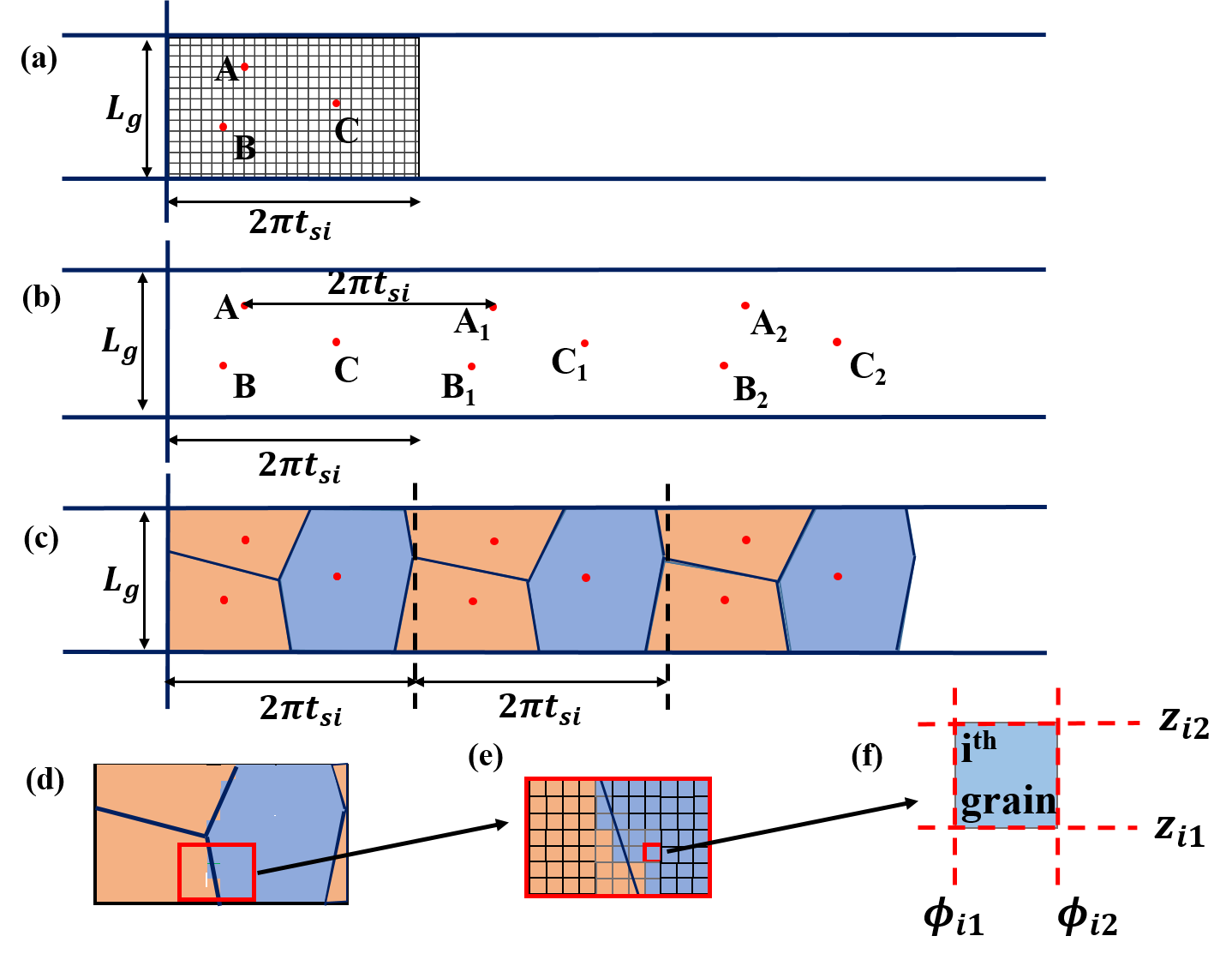}
\caption{Methodology to assign boundary conditions on Gate. (a) Gate is
discretized and using Poisson distribution, integers are generated randomly
and assigned to every grid point. Non-zero integer locations are
highlighted with red dots. (b) For each grain center, a point is marked at a
distance of $2\pi t_{si}$ and another at $2 \times 2 \pi t_{si}$ and so on. (c) The blue solid lines represent perpendicular bisectors for the lines joining the dots. For
each grain, WF is assigned based on probability. See the color assigned to
each grain; (d) the marked section in figure (c) is selected as gate area. This
selected area satisfies periodic boundary condition. (e) A representation of
discretization a large Voronoi grain into tiny squares of appropriate WF; (f)
The spatial coordinates ($z_{i1}$, $z_{i2}$, $\phi_{i1}$, $\phi_{i2}$) of $i$\textsuperscript{th} grid point are shown.}
\label{fig:methodology}
\end{figure}
\item For each grain center (say) ‘$A$’ (see Fig. 3(b)), a grain center at a horizontal distance of $2\pi t_{si}$ is marked as $A_1$ and another at a distance of $2 \times 2\pi t_{si}$ marked as $A_2$ and so on. The same procedure is followed for other grain centers $B$ and $C$. (For the sake of simplicity only three grain centers are assumed in Fig. 3(a)). Thus, periodicity among grain centers is maintained.
\item The model utilizes the Voronoi algorithm to create grain-boundaries and to divide the area into randomly shaped grains. The Voronoi algorithm produces realistic grain shapes \cite{ref12} and is explained in \cite{ref8}.
\item The section marked by the black dashed lines in Fig.3(c) is then considered as the gate area to be wrapped around the NWFET. This procedure ensures periodic boundary condition for the potential of the gate area as seen in Fig. 3(d).
\item The $i$\textsuperscript{th} grid-point corresponds to a square domain which lies between $[\phi_{i1},\phi_{i2}]$ and $[z_{i1},z_{i2}]$. This domain is given the appropriate WF value (Fig. 3(e-f)) to make grains on NWFET as shown in Fig. 1(a).
\end{enumerate}
Due to MGG, the WF vary randomly on the gate surface of NWFET. Depending on the material chosen for gate metal, the WF can take multiple values. For this work, TiN is assumed to be the gate metal.
\subsection{Computing the Electrostatic Potential in the Nanowire}
For solving the Laplace equation, the following boundary conditions for the cylinder are given below.

For the Gate:
\begin{equation}
V\vert_{r=t_{si}}=V_{GS} + \Phi_{MS}(\phi,z)
\end{equation}
For the Source:
\begin{equation}
V\vert_{y=0}=\Phi_{SC}
\end{equation}
For the Drain:
\begin{equation}
V\vert_{y=L_g}=\Phi_{SC} + V_{DS}
\end{equation}
where $V_{GS}$ is the applied gate bias, $\Phi _{MS}(\phi,z)$ is the work-function (spatially varying) difference between gate material (TiN) and channel. $\Phi_{SC}$ is the work-function difference between the Source (or the Drain ) and channel and $V_{DS}$ is the applied Drain bias.

The potential ( $V(r,\phi, z)$) inside the nanowire requires contributions from the Gate, Source and, Drain. The potential due to Gate ($V_G$) is computed using \eqref{eq:gatepotential} in MATLAB\textsuperscript{TM} whose derivation is adopted from \cite{jackson1975electrodynamics} which solves Laplace equation in cylindrical coordinates unlike Cartesian coordinate in \cite{ref8}. In the same way, the contribution of the Drain and the Source are computed using Eq. \eqref{eq:drainpotential} and Eq. \eqref{eq:sourcepotential} adopted from [13]. The Fourier Bessel coefficients for the Gate are computed using \eqref{eq:fbcoeff1} and \eqref{eq:fbcoeff2}. The total potential inside the nanowire at any point is the sum of all potentials from the Gate, the Source and, the Drain as given by equation \eqref{eq:superposition} by simply following the Superposition principle.
\begin{equation}\label{eq:gatepotential}
V_G(r,\phi,z)=\frac{1}{\pi L_g}\sum_{m=0}^{\infty}\sum_{n=1}^{\infty}\left[\frac{I_m(\frac{n\pi r}{L_g})}{I_m(\frac{n\pi t_{si}}{L_g})}\sin(\frac{n\pi z}{L_g})[C_{m,n}e^{i m\phi} + D_{m,n}e^{-i m\phi}]\right]
\end{equation}
\begin{equation}\label{eq:drainpotential}
V_D(r,\phi,z)=\sum_{m=0}^\infty 2(V_{DS} +\Phi_{SC})\frac{\sinh(\frac{X_m z}{t_{si}})}{\sinh(\frac{X_m L_g}{t_{si}})}\frac{J_0 (\frac{X_m r}{t_{si}})}{X_m J_1(X_m)}
\end{equation}
\begin{equation}\label{eq:sourcepotential}
V_S(r,\phi,z)=\sum_{m=0}^\infty 2\Phi_{SC}\frac{\sinh(\frac{X_m (L_g -z)}{t_{si}})}{\sinh(\frac{X_m L_g}{t_{si}})}\frac{J_0 (\frac{X_m r}{t_{si}})}{X_m J_1(X_m)}
\end{equation}
\begin{equation}\label{eq:fbcoeff1}
C_{m,n}=\sum_{p=1}^N \left[ \int_{zp1}^{zp2}\int_{\phi p1}^{\phi p2} V(r=t_{si}, z, \phi) \sin(\frac{n\pi z}{L_g})e^{-im\phi} d\phi dz \right]
\end{equation}
\begin{equation}\label{eq:fbcoeff2}
D_{m,n}=C_{m,n}^\dagger
\end{equation}
\begin{equation}\label{eq:superposition}
V(r,\phi,z)=V_G + V_D +V_S
\end{equation}
where $I_m$ is the Modified Bessel function of $m^{th}$ order, $J_0$ and $J_1$ are the Bessel functions of zeroth and first order respectively. $X_m$ is the $m^{th}$ zero of the zeroth order Bessel function, $C_{m,n}$ and $D_{m,n}$ are the Fourier Bessel coefficients and are complex conjugates to each other as shown in \eqref{eq:fbcoeff2}. $N$ is the total number of grains on the gate area of a single device. $V((r=t_{si}),\phi,z)$ is the boundary condition on the gate which varies spatially due to MGG.
\subsection{Percolation model to estimate $V_T$}
$V_T$ estimation by resistance based percolation model is adopted from \cite{ref14} in which local resistivity is computed from local potential $V(r,\phi,z)$ by Eq. \eqref{eq:percolation}. The effective resistance is given by Eq. \eqref{eq:Rtotal}. This model is same as the percolation model used for FinFET in \cite{ref8} except for geometry. The effective resistance is computed for each value (at a few regular intervals) of applied gate potential $V_{GS}$. We get exponentially decreasing $R_{total}$ values (and hence exponentially increasing current $I_D$ values computed using Eq. \eqref{eq:draincurrent} ) as we increase $V_{GS}$ values.
\begin{equation}\label{eq:percolation}
\rho(r,\phi,z)=\frac{1}{q\mu n_{S/D}}e^{\frac{qV_{BH}(r,\phi,z,V_{GS})}{kT}}=\frac{1}{q\mu n_{S/D}}e^{\frac{q(\Phi_{SC}-V(r,\phi,z,V_{GS}))}{kT}}
\end{equation}
\begin{equation}\label{eq:Rtotal}
R_{total}=\int_0^{L_g} \frac{dz}{\int_0^{t_{si}} \int_0^{2\pi} \frac{r dr d\phi}{\rho(r,\phi,z)}} 
\end{equation}
where $V_{BH}$ is the potential at the Source end, $n_{S/D}$ is the doping in the Source and Drain region.
\begin{equation}\label{eq:draincurrent}
I_D=\frac{V_{DS}}{R_{total}}
\end{equation}
For each device, the sub-threshold I-V is estimated and hence, $V_T$ (by constant current method) value is extracted. The same process is followed and repeated for many devices (e.g. 250 NWFETs) to get $V_T$ values and the standard deviation ($\sigma(V_T)$) for the distribution of values can be computed.

\section{Validation of the Proposed Model}
The analytical model is corroborated with 3D TCAD simulations performed using Sentaurus. A sample structure used for stochastic simulations in TCAD is shown in Fig.1(a). The TCAD simulation deck uses Density Gradient Model \cite{sdevicemanual} to account for quantization effects. The device and simulation parameters are calibrated according to the existing literature \cite{ref16,ref17}. First, electrostatic potentials are validated for few cases. Second, $V_T$ distributions obtained from analytical model are compared and validated against TCAD for different grain sizes. Finally, the standard deviations in $V_T$ distributions are compared to that of TCAD.
\begin{table}
\begin{tabular}[h]{|c|c|} 
 \hline
 \textbf{Parameter} & \textbf{Value}  \\ 
 \hline
 Physical Gate Length ($L_g$) & $20~nm$  \\ 
 \hline
 Nanowire diameter ($d=2r$) & $6~nm$  \\ 
 \hline
 Equivalent oxide thickness ($t_{ox}$) & $0.7~nm$ \\
 \hline
 Channel Doping & $10^{17}$ $cm^{-3}$ \\
 \hline
 S/D Doping & $10^{20}$ $cm^{-3}$ \\
 \hline
 WF for TiN for $<100>$ and $<111>$ respectively \\ 
 with probabilities 0.6 and 0.4 \cite{ref9}  & $4.6 ~ eV$ and $4.4 ~ eV$ \\
 \hline

\end{tabular}
\caption{Important parameters of the NWFET assumed in this work}
\end{table}
\subsection{Electrostatic potential validation with TCAD}

In this section, the potential (cylindrical Laplace solution) obtained from the model is compared with TCAD 3D simulations (for $V_{GS} =0 V$ and $V_{DS} = 0 V$). First, an NWFET with uniform gate WF (as shown in Fig. 4 (a)) displays good electrostatic potential agreement with TCAD , as seen in Fig. 4(c). Similarly, we show potential match for the non-uniform WF case (for the schematic of asymmetric WF grains as shown in Fig. 4(b)) in Fig. 4(d). 
\begin{figure}[h]
\centering
\includegraphics[scale=0.25]{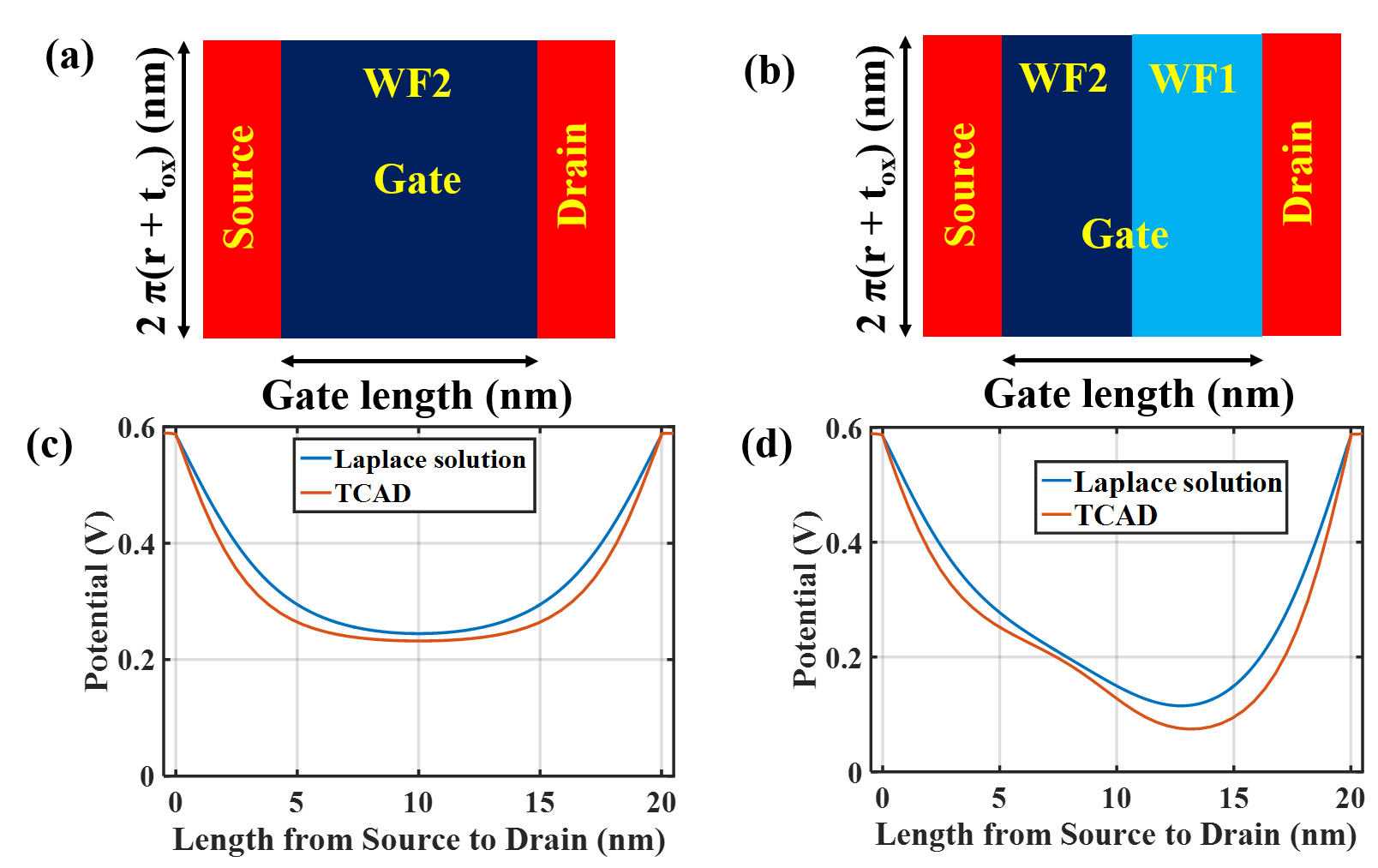}
\caption{Cut-open view of the gate of the NWFET (a) uniform WF; (b) non-
uniform WF (i.e. 2 grains); Comparison of electrostatic potentials between
TCAD and Laplace solutions at the center of the nanowire along the
channel length (at $r=0$) shows qualitative match for (c) uniform WF case;
and (d) non-uniform WF case.}
\label{fig:potentialvalidation}
\end{figure}
The asymmetry in WF (through the length of the device) is reflected in the asymmetry in the potential, which indicates that the positional dependence of the metal grains is accounted for. There seems to be a small quantitative mismatch in the potential. The gate oxide replacement, based on long cylinder approximation, is the primary reason for the mismatch in potential (near to Source or Drain, the tangential electric field dominates over the normal electric field). Along with that, use of finite basis functions in computation and assumption of metallic S/D also causes a mismatch in potential. Even so, the analytical Laplace solution is capable of capturing changes in the potential due to MGG qualitatively as shown in Fig. 4(d). This is essential to model $V_T$ for any arbitrary grain sizes, shape
and distribution. The small quantitative errors in absolute potential produce a mean shift in $V_T$ distribution. Potential estimated using cylindrical Laplace solution is of decent accuracy by taking into account only first few harmonics of the sinusoidal functions (around 10 in number) and Modified Bessel functions (around 5 in number) for NWFET parameters given in Table 1.
\subsection{Quantization effects on MGG induced $V_T$ variability} 
The quantum mechanical confinement effects due to narrow diameters in nanowires are not accounted for in our model. So, it is essential to have an estimate of error, if quantization is neglected. In TCAD, typically the Density Gradient (DG) model is activated in device simulations to account for this \cite{sdevicemanual}. We compare the Classical Drift-Diffusion (DD) with DG corrected DD (DG + DD) stochastic 3D TCAD $V_T$ distributions of 5 nm and 20 nm grain sizes (Fig. 5). 
\begin{figure}[h]
\centering
\includegraphics[scale=0.25]{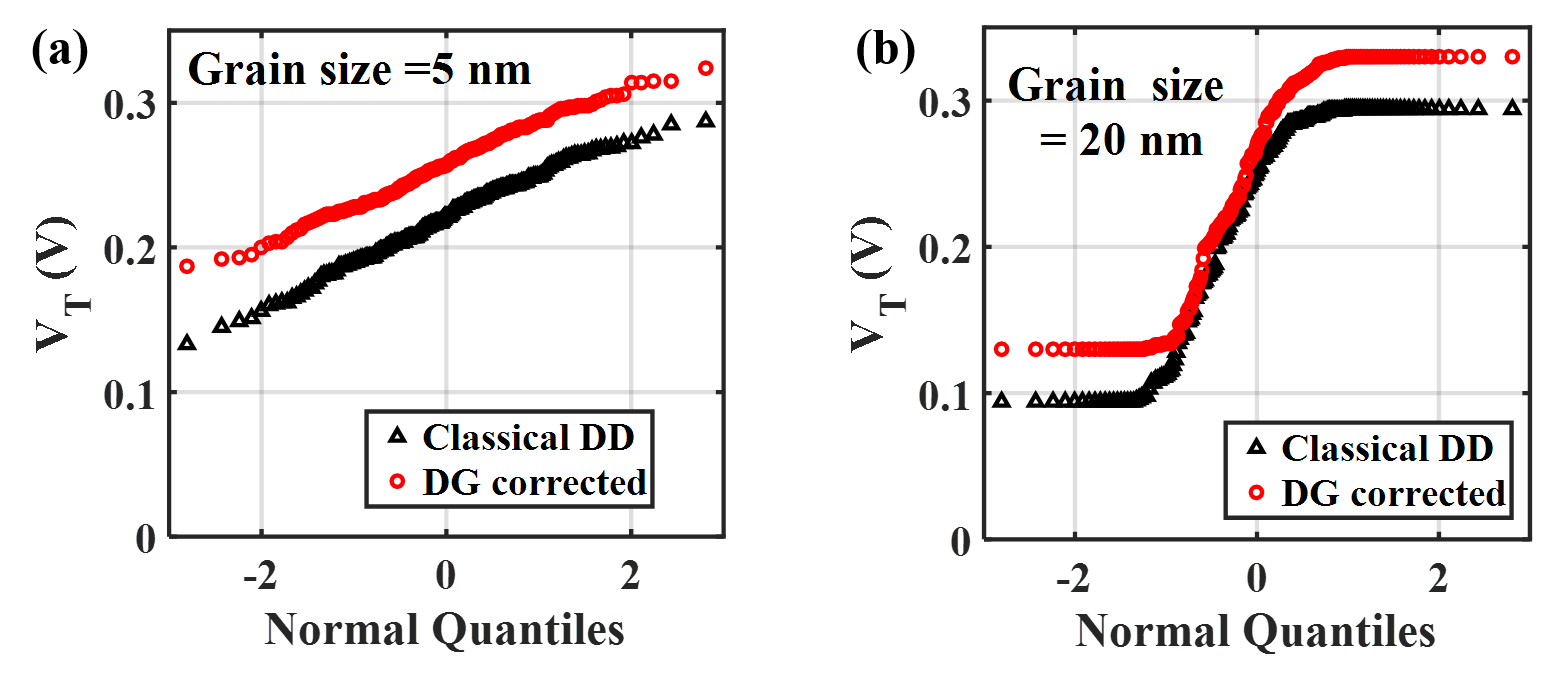}
\caption{Quantile plots of $V_T$ extracted using classical Drift-Diffusion (DD)
vs Density Gradient (DG) Correction in TCAD for an average grain size
of (a) 5 nm; (b) 20 nm. The distribution is simply shifted i.e. mean ($\mu$) is
shifted keeping the standard deviation ($\sigma$) intact.}
\label{fig:DGcomp}
\end{figure}
The $V_T$ distributions look identical except for the mean shift. From the Fig. 6(a), it is evident that standard deviation ($\sigma(V_T)$) is overestimated by 2 − 3 mV in the absence DG correction. The mean ($\mu$) however 20 − 30 mV shows a significant shift in $V_T$ distribution, which is expected \cite{ref18}. This shows that, quantization shows a small effect on MGG induced $V_T$ variability as shown in Fig. 6(a), i.e. essentially an offset, without affecting the shape of the distribution (see Fig. 5).

We also present the statistics of sub-threshold slope (SS). The mean sub-threshold slope ($\mu(SS)$) values of DG corrected distributions are slightly (negligibly) higher (worse) compared to classical case. Both DD and DG+DD standard deviations ($\sigma(SS)$) are much smaller compared to their respective mean values as shown in Fig. 6(b). Hence, the change in $V_T$ channeled via change in SS due to inclusion of DG is insignificant (see Fig. 6(c)). This shows that MGG either with or without DG correction has little impact on SS in NWFET, consistent with \cite{ref3}. 
\begin{figure}[h]
\centering
\includegraphics[scale=0.27]{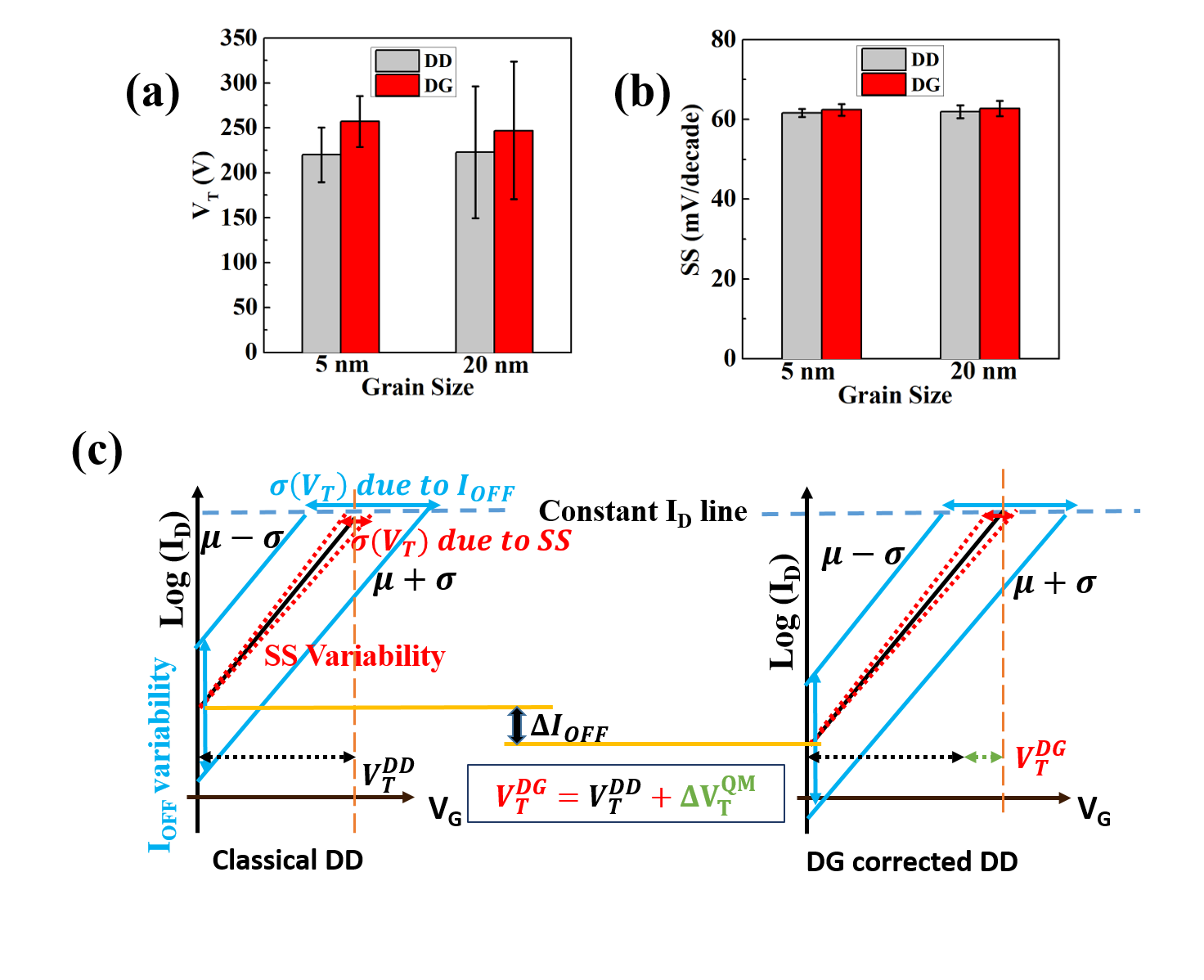}
\caption{(a) Bar plot of the $V_T$ shows that the DG simulations mean is shifted
without change in the standard deviation compared to Classical DD. (b)
Bar plot of the SS shows that the DG simulations mean is shifted by a very
small amount. (c) Schematic of the IV characteristics for $\mu$ and $\mu \pm \sigma$
showing the comparison between DD and DG. The fractional variation in
SS i.e. $\frac{\sigma}{\mu}\leq 2.5 \%$ is much smaller than $\frac{\sigma}{\mu}$ of the $V_T$ distribution, which is 13-26\%. This implies that the $V_T$ variability is ascribed to primarily initial electrostatic barrier at $V_G = 0$ (see $\Delta I_{OFF}$ in Fig. (c)), while gate-barrier modulation (represented by SS) is largely unaffected by MGG. Again, DG correction simply shifts the $\mu$, without affecting the $\sigma$ of the $V_T$ distribution.}
\label{fig:DGcomp2}
\end{figure}
In comparison with FinFET reported in \cite{ref8}, the quantization effects on MGG induced $V_T$ variability is slightly higher for NWFET. In order to account for quantization effects in NWFET, a more elaborate method or model is necessary to introduce corrections in potential and charge density, which is beyond the scope of this work.
\subsection{Comparison of $V_T$ distributions}
First, stochastic 3D simulations of NWFETs with MGG are performed in Sentaurus\textsuperscript{TM} TCAD. The TCAD $V_T$ distributions are then compared with our model for four average grain sizes (3, 5, 7, 10, 15 and 20 nm) in Fig. 7.
\begin{figure}[h]
\centering
\includegraphics[scale=0.22]{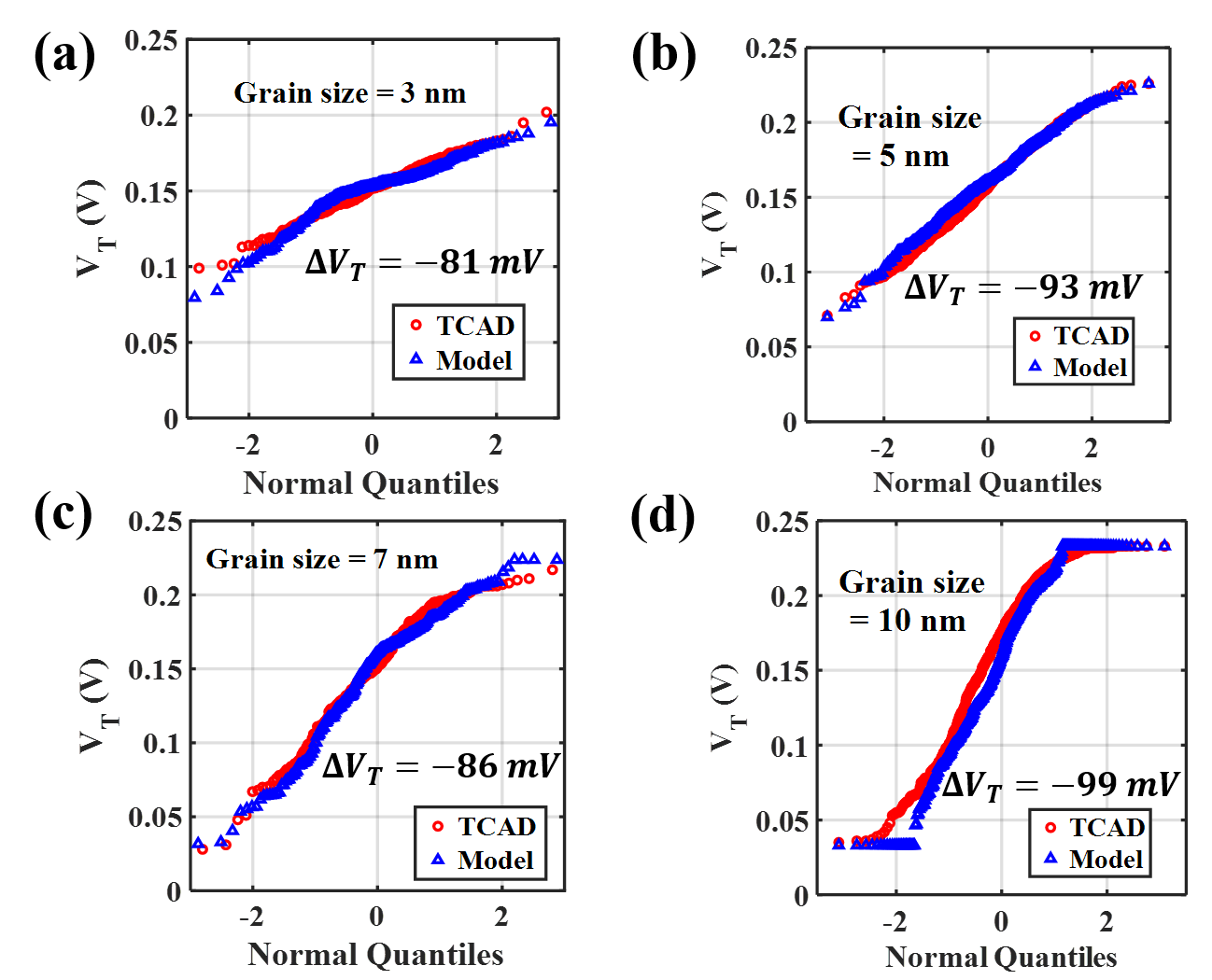}
\includegraphics[scale=0.225]{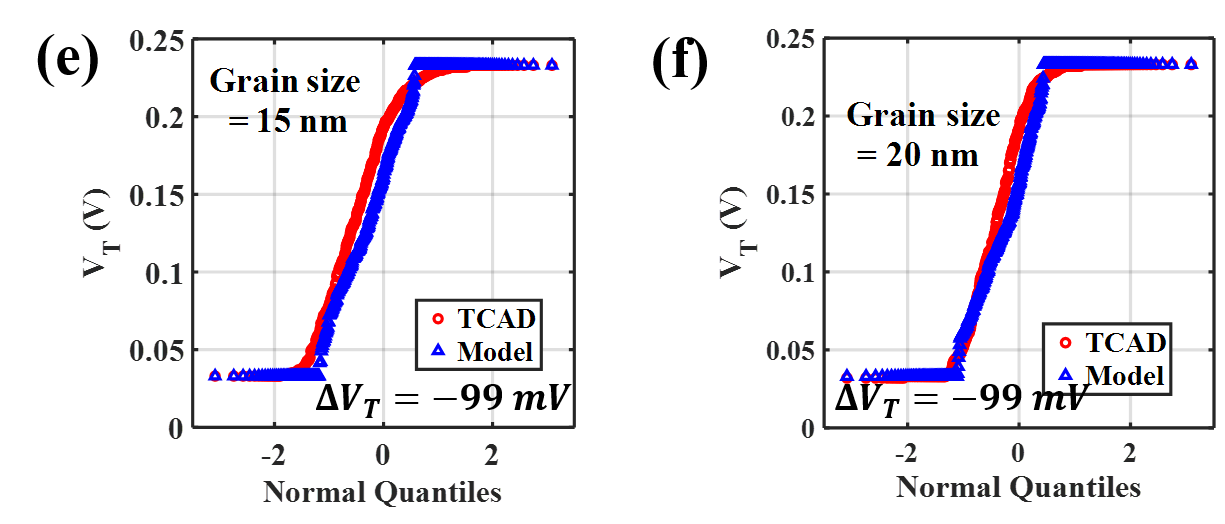}
\caption{QQ plots of V T distribution estimated by analytical model (blue triangles) is compared to TCAD simulations (red circles) for 250 samples for each grain size) for average grain size of (a) 3 nm, (b) 5 nm, (c) 7 nm, (d) 10 nm, (e) 15 nm, and (f) 20nm. A $V_T$ offset equal to the shift in $\mu$ is added to analytical model’s $V_T$ distribution to align with TCAD $V_T$ distribution to observe excellent quantitative match in distribution.}
\label{fig:qqplots}
\end{figure}
The mean ($\mu$) values of the $V_T$ distributions obtained from TCAD and proposed analytical model are different. So, for each $V_T$ distribution, an offset ($\Delta V_T$) is added to match the range of TCAD distribution. This alignment helps in better comparison of $V_T$ distributions with TCAD which can be observed in Fig. 7.

These simulations are performed only for $V_{DS} = 50mV$. This is based on earlier TCAD simulation study on MGG in NWFET devices in \cite{ref3} showed that the dependence of $\sigma(V_T)$ on $V_{DS}$ is negligible.

As revealed by Fig. 7, the quantile plot resembles a Gaussian distribution for small grain sizes. However, for larger grain sizes ($\geq 10 nm$), the $V_T$ distributions are Gaussian but truncated by the extreme $V_T$ values. The extreme values differ by 200 mV which is equal to the WF differences in TiN. In other words, for large grains sizes, single WF gated devices (of either orientations) are more probable which is already reported in \cite{ref3}.The proposed analytical model can capture the shape of the distribution effectively for wide range of grain sizes i.e. Gaussian like distribution for smaller grain sizes and non-
Gaussian for larger grain sizes.

\subsection{Comparison of standard deviations of $V_T$ distributions}
\begin{figure}[h]
\centering
\includegraphics[scale=0.28]{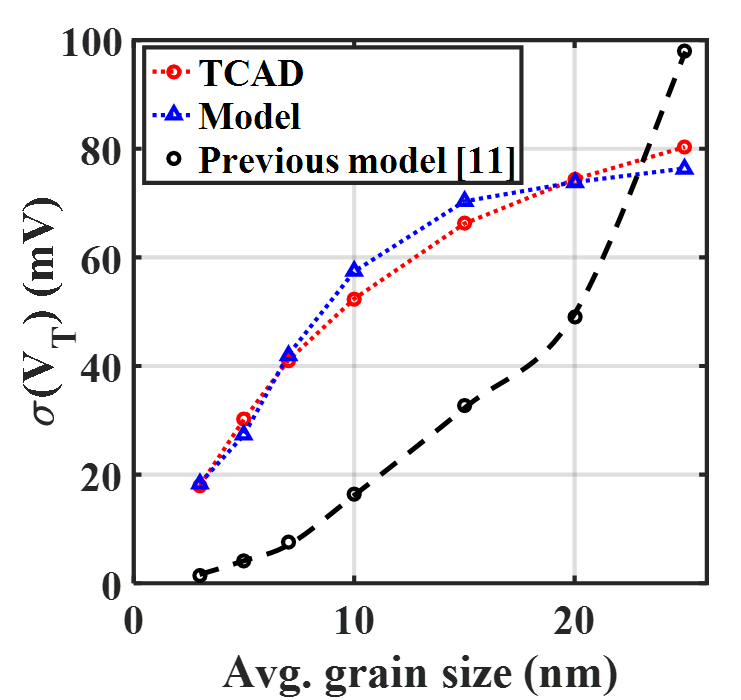}
\caption{Plot of standard deviations of $V_T$ values vs grain size. The size
dependence of $\sigma(V_T)$ is captured well by the analytical model (blue dotted
line with triangles) as compared to stochastic TCAD simulations (red
dotted line with circles). They show quantitative agreement with a 6\% error
(3.5mV RMS error over the average range of 51mV of $\sigma(V_T)$). In comparison, the
state-of-the-art analytical model (black dashed line with circles) from
literature \cite{ref9}, which is unable to qualitatively capture the trends. 250
devices for each grain size are considered.}
\label{fig:sigmavt}
\end{figure}
In this section, the $\sigma(V_T)$s extracted from the analytical model are compared to TCAD 3D stochastic simulations (250 devices for every grain size). Both qualitative and quantitative match is observed in Fig. 8 between the model and the TCAD.
For comparison, we also plotted the $\sigma(V_T)$ obtained using analytical model proposed earlier in \cite{ref9}. The proposed analytical model gives about $3.5mV$ RMS error when compared with TCAD results. The relative error is small i.e. 6$\%$ compared to mean of $\sigma(V_T) \approx 51mV$ over the range of grain sizes. Our analytical method is significantly ($63 \times$) faster than TCAD ( On a computer with Intel Xeon\textsuperscript{TM} CPU E5620 (2.4 GHz × 16) processor with 64 GB of RAM).Thus, the proposed analytical model is reasonably accurate in $\sigma(V_T)$ prediction with high efficiency in computation.

\section{Conclusion}
MGG induced $V_T$ variability in NWFETs is estimated using electrostatics (along with percolation) based analytical modeling. The model successfully adapts the FinFET MGG model to cylindrical NWFET geometry with three modifications. They are 1) adoption of periodic boundary condition of gate potential in the azimuthal direction; 2) used long cylinder approximation to replace oxide with Silicon; 3)
solved Laplace equations in cylindrical coordinate system. The $V_T$ distributions, and $\sigma(V_T)$ are validated with stochastic 3D TCAD simulations for different grain sizes. The effects of quantization although expected to be present at these dimensions, produces a shift in the $V_T$ distribution without affecting its shape. Thus, $\sigma(V_T)$ also remains unaffected. This model accurately captures the $V_T$ distribution for a large range of grain sizes (3-20 nm) – validated against 3D TCAD simulations. The $\sigma(V_T)$ dependence on grain size is quantitatively accurate, which is a significant improvement over the state-of-the-art, which is unable to capture qualitative trends. The proposed model has 63 × lesser computational cost when compared to stochastic 3D TCAD simulations. Such a model is able to capture the non-Gaussian $V_T$ distribution to enable accurate circuit simulation for device-circuit interactions.

\section*{References}

\bibliography{sample.bib}

\begin{thebibliography}{10}
\expandafter\ifx\csname url\endcsname\relax
  \def\url#1{\texttt{#1}}\fi
\expandafter\ifx\csname urlprefix\endcsname\relax\def\urlprefix{URL }\fi
\expandafter\ifx\csname href\endcsname\relax
  \def\href#1#2{#2} \def\path#1{#1}\fi

\bibitem{ref1}
K.~J. Kuhn, Considerations for ultimate cmos scaling, IEEE Transactions on
  Electron Devices 59~(7) (2012) 1813--1828.
\newblock \href {http://dx.doi.org/10.1109/TED.2012.2193129}
  {\path{doi:10.1109/TED.2012.2193129}}.

\bibitem{ref2}
X.~Wang, A.~R. Brown, B.~Cheng, A.~Asenov, Statistical variability and
  reliability in nanoscale finfets, in: 2011 International Electron Devices
  Meeting, 2011, pp. 5.4.1--5.4.4.
\newblock \href {http://dx.doi.org/10.1109/IEDM.2011.6131494}
  {\path{doi:10.1109/IEDM.2011.6131494}}.

\bibitem{ref3}
K.~Nayak, S.~Agarwal, M.~Bajaj, P.~J. Oldiges, K.~V. R.~M. Murali, V.~R. Rao,
  Metal-gate granularity-induced threshold voltage variability and mismatch in
  si gate-all-around nanowire n-mosfets, IEEE Transactions on Electron Devices
  61~(11) (2014) 3892--3895.
\newblock \href {http://dx.doi.org/10.1109/TED.2014.2351401}
  {\path{doi:10.1109/TED.2014.2351401}}.

\bibitem{ref4}
N.~Seoane, G.~Indalecio, M.~Aldegunde, D.~Nagy, M.~A. Elmessary, A.~J.
  García-Loureiro, K.~Kalna, Comparison of fin-edge roughness and metal grain
  work function variability in ingaas and si finfets, IEEE Transactions on
  Electron Devices 63~(3) (2016) 1209--1216.
\newblock \href {http://dx.doi.org/10.1109/TED.2016.2516921}
  {\path{doi:10.1109/TED.2016.2516921}}.

\bibitem{ref5}
X.~Jiang, X.~Wang, R.~Wang, B.~Cheng, A.~Asenov, R.~Huang, Predictive compact
  modeling of random variations in finfet technology for 16/14nm node and
  beyond, in: 2015 IEEE International Electron Devices Meeting (IEDM), 2015,
  pp. 28.3.1--28.3.4.
\newblock \href {http://dx.doi.org/10.1109/IEDM.2015.7409787}
  {\path{doi:10.1109/IEDM.2015.7409787}}.

\bibitem{ref6}
H.~Dadgour, K.~Endo, V.~De, K.~Banerjee, Modeling and analysis of
  grain-orientation effects in emerging metal-gate devices and implications for
  sram reliability, in: 2008 IEEE International Electron Devices Meeting, 2008,
  pp. 1--4.
\newblock \href {http://dx.doi.org/10.1109/IEDM.2008.4796792}
  {\path{doi:10.1109/IEDM.2008.4796792}}.

\bibitem{ref7}
H.~Nam, Y.~Lee, J.~D. Park, C.~Shin, Study of work-function variation in high-
  $\kappa $ /metal-gate gate-all-around nanowire mosfet, IEEE Transactions on
  Electron Devices 63~(8) (2016) 3338--3341.
\newblock \href {http://dx.doi.org/10.1109/TED.2016.2574328}
  {\path{doi:10.1109/TED.2016.2574328}}.

\bibitem{ref8}
P.~H. Vardhan, S.~Mittal, S.~Ganguly, U.~Ganguly, Analytical estimation of
  threshold voltage variability by metal gate granularity in finfet, IEEE
  Transactions on Electron Devices 64~(8) (2017) 3071--3076.
\newblock \href {http://dx.doi.org/10.1109/TED.2017.2712763}
  {\path{doi:10.1109/TED.2017.2712763}}.

\bibitem{ref10}
B.~Ray, S.~Mahapatra, Modeling and analysis of body potential of cylindrical
  gate-all-around nanowire transistor, IEEE Transactions on Electron Devices
  55~(9) (2008) 2409--2416.
\newblock \href {http://dx.doi.org/10.1109/TED.2008.927669}
  {\path{doi:10.1109/TED.2008.927669}}.

\bibitem{ref11}
P.~H. Vardhan, S.~Mittal, A.~S. Shekhawat, S.~Ganguly, U.~Ganguly, Analytical
  modeling of metal gate granularity induced vt variability in nwfets, in: 2016
  74th Annual Device Research Conference (DRC), 2016, pp. 1--2.
\newblock \href {http://dx.doi.org/10.1109/DRC.2016.7548446}
  {\path{doi:10.1109/DRC.2016.7548446}}.

\bibitem{ref9}
H.~F. Dadgour, K.~Endo, V.~K. De, K.~Banerjee, Grain-orientation induced work
  function variation in nanoscale metal-gate transistors; part i: Modeling,
  analysis, and experimental validation, IEEE Transactions on Electron Devices
  57~(10) (2010) 2504--2514.
\newblock \href {http://dx.doi.org/10.1109/TED.2010.2063191}
  {\path{doi:10.1109/TED.2010.2063191}}.

\bibitem{ref12}
S.~H. Chou, M.~L. Fan, P.~Su, Investigation and comparison of work function
  variation for finfet and utb soi devices using a voronoi approach, IEEE
  Transactions on Electron Devices 60~(4) (2013) 1485--1489.
\newblock \href {http://dx.doi.org/10.1109/TED.2013.2248087}
  {\path{doi:10.1109/TED.2013.2248087}}.

\bibitem{jackson1975electrodynamics}
J.~D. Jackson, Electrodynamics, Wiley Online Library, 1975.

\bibitem{ref14}
S.~Mittal, A.~S. Shekhawat, U.~Ganguly, An analytical model to estimate
  finfet's $v_{T}$ distribution due to fin-edge roughness, IEEE Transactions on
  Electron Devices 63~(3) (2016) 1352--1358.
\newblock \href {http://dx.doi.org/10.1109/TED.2016.2520954}
  {\path{doi:10.1109/TED.2016.2520954}}.

\bibitem{sdevicemanual}
Synopsys, Sentaurus Device User Guide, Synopsys Inc, 2015.

\bibitem{ref16}
S.~Mittal, S.~Gupta, A.~Nainani, M.~C. Abraham, K.~Schuegraf, S.~Lodha,
  U.~Ganguly, Epitaxially defined finfet: Variability resistant and
  high-performance technology, IEEE Transactions on Electron Devices 61~(8)
  (2014) 2711--2718.

\bibitem{ref17}
N.~Pons, F.~Triozon, M.-A. Jaud, R.~Coquand, S.~Martinie, O.~Rozeau, Y.-M.
  Niquet, V.-H. Nguyen, A.~Idrissi-El~Oudrhiri, S.~Barraud, Density gradient
  calibration for 2d quantum confinement: Tri-gate soi transistor application,
  in: Simulation of Semiconductor Processes and Devices (SISPAD), 2013
  International Conference on, IEEE, 2013, pp. 184--187.

\bibitem{ref18}
H.~Mehta, S.~Lodha, U.~Ganguly, S.~Ganguly, Calibration of the density-gradient
  tcad model for germanium finfets, in: Micro and Nanoelectronics (RSM), 2013
  IEEE Regional Symposium on, IEEE, 2013, pp. 143--146.

\end{thebibliography}

\end{document}